\documentstyle[anta, twoside]{article}

\input{psfig.sty}

\def\Journal#1#2#3#4{(#1) {#2} {\bf #3}, #4}

\def\AJ{\em Astron.~J.}

\def\AaApS{\em Astron. Astrophys. Suppl.}

\def\MNRAS{\em Mon. Not. R.~Astron. Soc.}

\newcommand{\HI}{{\rm H\,\scriptstyle I}}

\newcommand{\degs}{{$^{\circ}$}}
\begin{document}

\markboth{M. Wolleben et al.}{New Northern Sky Polarization Survey}

\thispagestyle{plain}
\setcounter{page}{51}

\title{The DRAO 26-m Large Scale Polarization Survey\\at 1.41 GHz}

\author{M. Wolleben$^{1,2}$, T.L. Landecker$^2$, W. Reich$^1$ and R. Wielebinski$^1$}

\address{$^1$Max-Planck-Institut f\"ur Radioastronomie,
              Auf dem H\"ugel 69, 53121 Bonn, Germany\\
         $^2$Dominion Radio Astrophysical Observatory, Herzberg
             Institute of Astrophysics, National Research Council of
             Canada, Box 248, Penticton, BC V2A 6J9, Canada }

%%%%%%%%%%%%%%%%%%%%%%%%%%%%%%%%%%%%%%%%%%%%%%%%%%%%%%%%%%%%%%%%%%%%

\maketitle

\abstract{The Effelsberg telescope as well as the DRAO synthesis
telescope are currently surveying the Galactic polarized emission at
$\lambda\,21$~cm in detail. These new surveys reveal an unexpected
richness of small-scale structures in the polarized sky. However,
observations made with synthesis or single-dish telescopes are not on
absolute intensity scales and therefore lack information about the
large-scale distribution of polarized emission to a different degree.
Until now, absolutely calibrated polarization data from the
Leiden/Dwingeloo polarization surveys are used to recover the missing
spatial information. However, these surveys cannot meet the
requirements of the recent survey projects regarding sampling and
noise and new polarization observation were initiated to complement
the Leiden/Dwingeloo Survey. In this paper we will outline the
observation and report on the progress for a new polarization survey
of the northern sky with the 26-m telescope of the DRAO.}

\section{Motivation}
Over the past several years a number of survey projects were launched
in order to map the polarized emission from the Galaxy in great detail
(e.g. Junkes et al. 1987, Duncan et al. 1997, Uyan{\i}ker et al. 1998,
Taylor et al. 2003).
Single-dish as well as synthesis telescopes located in the northern
and southern hemisphere are being used. However, observations made
with synthesis telescopes are systematically affected by a missing
zero spacing problem, which prevents the detection of extended
emission. In a similar way, single-dish observations can lack a
large-scale emission component, which is often lost during the data
reduction process when attempts are made to subtract the ground
radiation from the observations. If the ground radiation is not known
exactly, one cannot distinguish between real sky signal and radiation
from the ground.

For an interpretation of polarimetric observations, an accurate
absolute calibration is essential. Other than for total power
observations, absolute calibration of polarization data means to
calibrate vectorial quantities. Surprisingly on first sight, this
means that absolute calibration of polarized intensity maps can turn
emission-type objects into minima in PI, and vice versa. Also rotation
measures can be affected by an inaccurate calibration of polarization
data.

Missing zero spacings in interferometric observations are added from
single-dish observations. In turn, large-scale components in
single-dish data are recovered by using existing absolutely calibrated
polarization data from the 60's and 70's made with single-dish
telescopes, e.g. the Leiden/Dwingeloo polarization surveys (Brouw \&
Spoelstra 1976). However, the severe under-sampling and relatively
high noise of these data often makes an accurate calibration of the
recent surveys impossible. Until now, these absolutely calibrated data
from the Leiden/Dwingeloo surveys laid the foundation for the
calibration chain up to the synthesis telescopes, and therefore an
improvement of these {\it old} data regarding sampling and noise
became necessary.

\section{The DRAO 26-m Telescope and Receiver}

The 26-m telescope at DRAO is on an equatorial mount, with a mesh
surface and the receiver placed in its prime focus on three feed-legs.
The pointing accuracy is about 48\arcsec. The receiver used for
polarimetric observations consists of a quadrature HF-hybrid which
forms circular out of linear polarization, two uncooled FET-amplifiers
for both hands of polarization operating on room temperature, and
circulators to improve matching. The system temperature was about
$150$~K. Two HF-bandpass filters and additional IF-filters were used
to limit the bandwidth to $12$~MHz ($10$~MHz for data obtained
in November 2002). An integrated noise source was used to inject a
calibration signal every $32$~seconds for a duration of $400$~ms into
the X and Y part of the receiving system. Gain variations were less
than $4$\% over a period of $40$~days.

The IF-polarimeter is an analog 2-channel multiplier providing the
four correlation products RR, LL, RL, and LR. The polarimeter was
brought from the MPIfR, Bonn and is of the type used on the Effelsberg
telescope. It operates at an IF of $150$~MHz and a total bandwidth of
$50$~MHz. Every $4$~seconds, an internal phase shifter was switched to
invert the signs of the two cross-correlation products. This was
necessary to correct for any electronic drifts in the polarimeter.
Data were integrated and recorded every $40$~ms.

\begin{figure}[tb]
\centerline{\psfig{figure=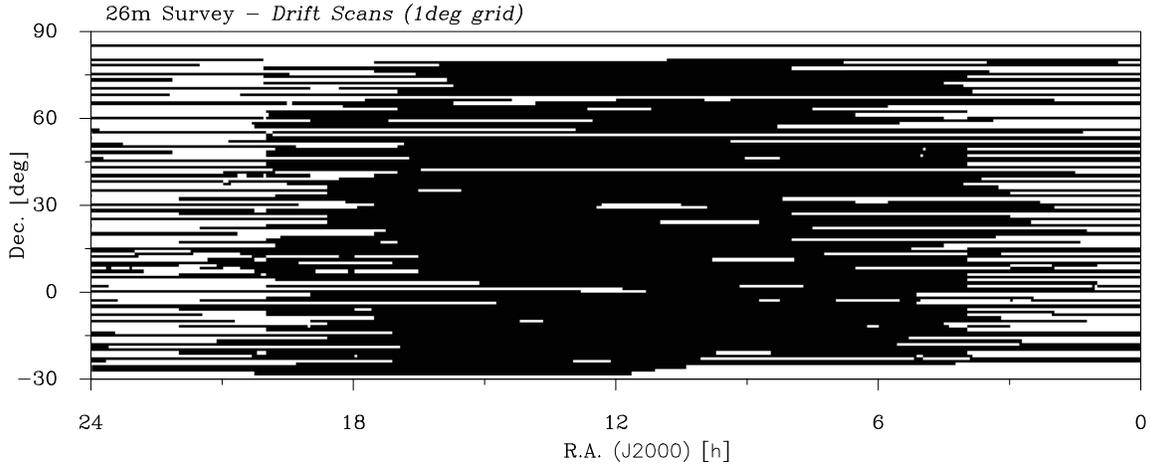,angle=270,width=15.5truecm}}
\caption{Map showing the sky coverage of the new survey. Black lines
indicate drift scans. }
\end{figure}

\section{Observation, Calibration and Reduction Steps}

Observations were made during seven months from November 2002 through
May 2003 using the 26-m telescope in the drift-scan mode. Each night
one declination was observed. We made 168 drift scans covering the
entire sky observable from the DRAO with the telescope stationary on
the meridian. Many regions could be observed with a $1$\degs~separation
in declination or smaller; some could only be observed with a
$5$\degs~sampling in declination. Along right ascension the drift
scans are fully sampled. The center frequency of the receiver was
tuned to $1410$~MHz, avoiding the $\HI$ line emission. Most of the
observations were done during the night. Prior and after each drift
scan, standard calibration sources were mapped.

The hybrid introduces cross-talk and extensive calibration was needed
to transform the four polarimeter output channels into true Strokes
parameters. The following steps were applied to the raw data:

\begin{description}

%flagging
\item[Flagging:]
Interference removal was done automatically by a flagging algorithm
which searches for short-term peaks which are
$4.5\,\sigma_{\mathrm{rms}}$ above the average signal level over a
$2$~minute time window.

% Mueller Matrix
\item[Pre-scaling and rotation:]
The vector formed out of the four correlation products RR, LL, RL, and
LR is multiplied with a $4\times4$--matrix, similar to a M\"uller
matrix with the difference that correlation products instead of Stokes
parameters are transformed. This reduction step includes a pre-scaling
and rotation in order to transform the arbritrary polarimeter units
into meaningful temperatures in the equatorial system. An iterative
method is used to find the appropriate correction matrix. This
algorithm starts with the unit matrix and successively modifies matrix
entries until the best--fit matrix is found. The quality of the matrix
can be estimated by the correlation of the corrected raw data with the
absolutely calibrated Leiden/Dwingeloo polarization data, resp. the
Stockert total-power data (Reich 1982, Reich \& Reich 1986).

% Receiver Gain and Phase drifts
\item[Gain and phase:]
Electronic gain and phase variations in the receiving system were
recorded by changes in the signal level of the calibration signal in
all four channels. The standard temperature and position angle of the
calibration signal assumed for this survey is given by the average
values over the seven months.

% Elevation Dependent Polarization
\item[Ground radiation:]
Spurious polarization is picked up by the side-lobes of the telescope.
This mainly ground-based signal can be intrinsically unpolarized,
however, the side-lobes are highly polarized, which results in an
elevation- and azimuth-dependent polarization component. Since we
exclusively observed in the meridian we only have to deal with the
elevation dependency. For the purpose of deriving a ground-radiation
model we made elevation scans along the meridian at different sidereal
times. Their average served as a time-independent model (see
Fig.~\ref{bildchen_ground}) which was used to subtract the
ground-radiation component.
\begin{figure}[b]
\centerline{\psfig{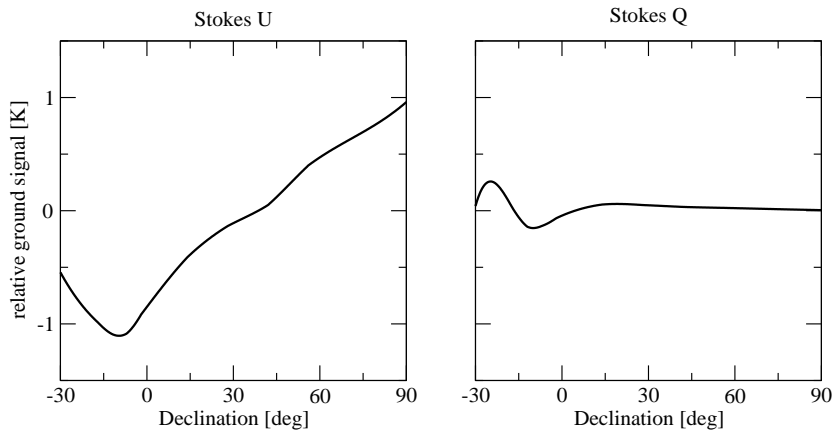}}
\caption{Ground radiation profiles for Stokes~$U$ and $Q$ derived by
averaging elevation scans. }
\label{bildchen_ground}
\end{figure}

% Ionospheric and Sun
\item[Sun and ionospheric Faraday rotation:]
Beside the rather stable ground radiation, which was modelled, the sun
is another strong source of spurious polarization. In the same way as
the ground is detected as a polarized signal, the bright signal from
the sun also causes strong polarized signals if radiating into the
side-lobes. Therefore observations were limited to night-time to
circumvent interference from the sun. In addition, ionospheric Faraday
rotation can affect the position angle of polarized signals. This
effect is negligible during the night. Whenever day-time data were
used, we cross-checked the data against neighboring drift scans for
possible sun interference or ionospheric Faraday rotation.

% De-striping
\item[De--striping:]
Although the receiving system is fairly stable, there are remaining
variations in the system temperature. We believe that these variations
are caused by seasonal changes of the ground radiation or temperature
drifts in the front-end part of the receiver. Changing offsets in the
system temperature show up as stripes in the final map and we compared
in an iterative method each drift scan with its neighboring scans to
correct for variations in the system temperature.

% Final rotation scaling and instr. pol 2nd matrix
\item[Final scaling, rotation, and instrumental polarization:]
After applying the reduction steps discussed above, a second pass
M\"uller matrix fit was necessary to improve the correlation with the
reference values from the Leiden/Dwingeloo survey. Also the cross-talk
introduced by the receiver was now corrected.

\end{description}

The digitized version of the Leiden/Dwingeloo data at $1.4$~GHz
provides Stokes~$U$ and $Q$ brightness temperatures for more than
$1600$ different positions on the sky. About $700$ positions are
covered by our survey within a radius of $15$\arcmin. These common
data points serve as reference values for the matrix fits discussed
above. This means that our temperature scale fully relies on the
Leiden/Dwingeloo data.

\section{Example Drift Scan}
% 56 deg declination
\begin{figure}[tb]
\centerline{\psfig{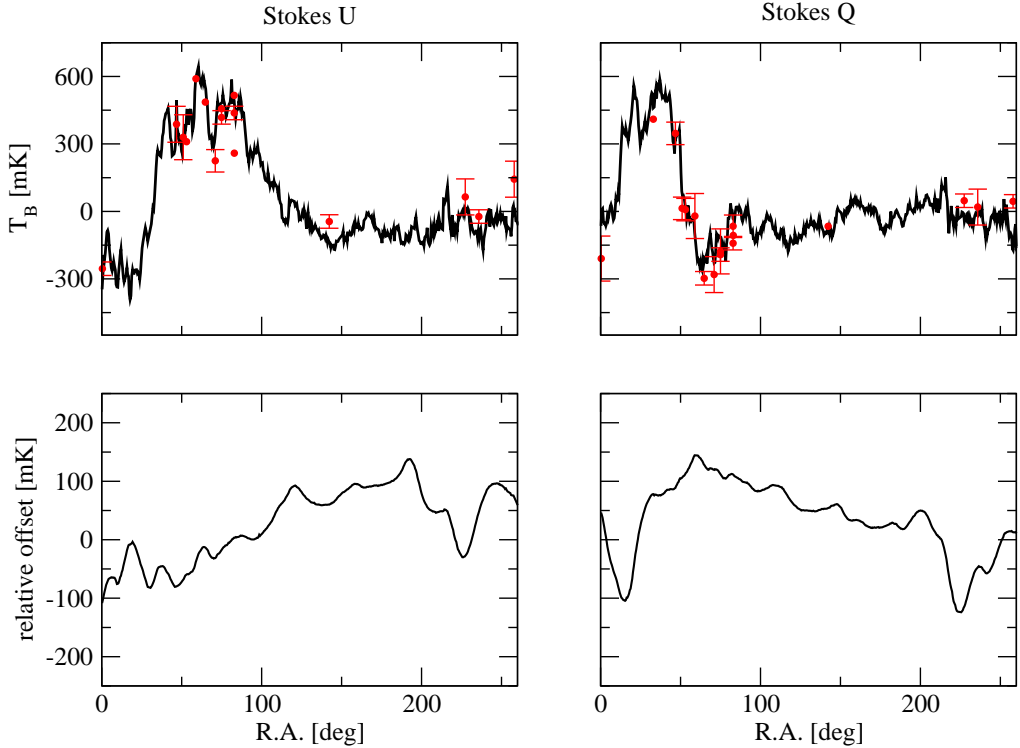}}
\caption{Drift scan at $56$\degs~declination. The upper panels show
Stokes~$U$ and $Q$ brightness temperatures. Available data points from
the Leiden/Dwingeloo survey are overlaid with error bars indicating
the quoted errors. The lower panels show the relative system
temperature derived from the de--striping algorithm.}
\label{bildchen_driftscan}
\end{figure}
In Fig.~\ref{bildchen_driftscan} we show an example drift scan at
56\degs~declination. The drift scan passes the {\it fan}-region, a
bright, polarized region centered at about $\alpha=50$\degs,
$\beta=65$\degs. Considering the mean errors ($60$~mK as quoted for
the Leiden/Dwingeloo data, and about $30$~mK for the new survey), the
Stokes~$U$ and $Q$ distributions of both surveys are in good agreement
for this drift scan. Remarkable is the appearance of small-scale
structure which cannot be seen in the Leiden/Dwingeloo data set
because of its coarse sampling. Fig.~\ref{bildchen_driftscan} also
shows the offsets in Stokes~$U$ and $Q$, which apparently suffer from
a temperature drift. This offset drift is most likely caused by
variations of the receiver temperature or local environmental
conditions and was derived from the de--striping algorithm.

\section{Repeatability of Drift Scans}
\begin{figure}[tbp]
\centerline{\psfig{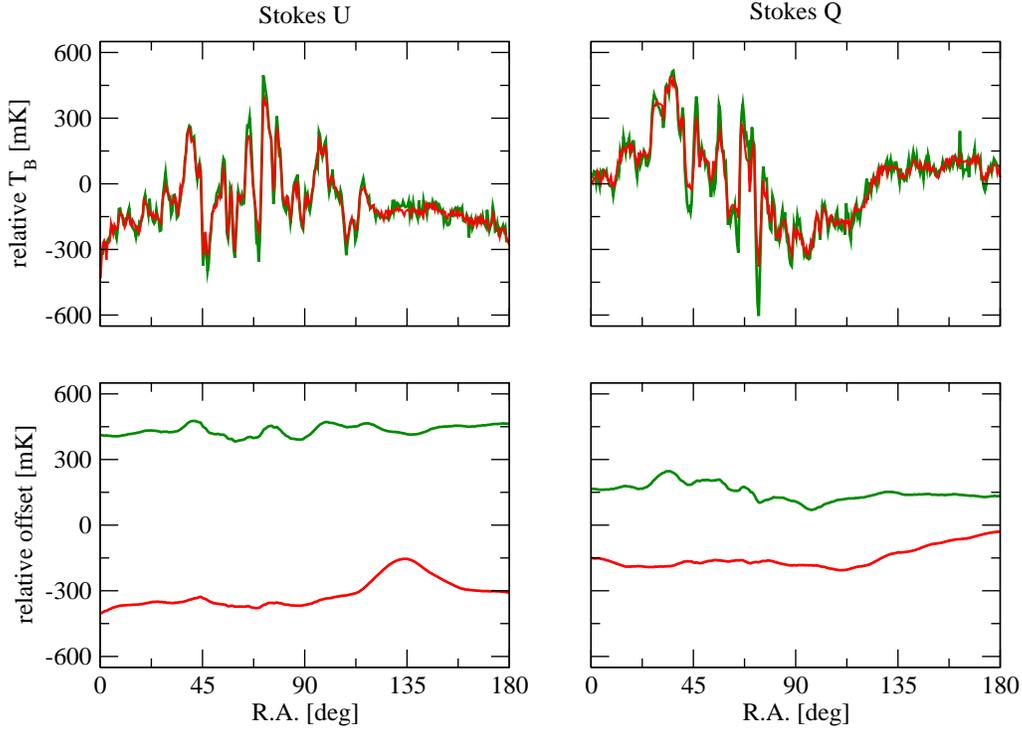}}
\caption{Two drift scans at $40$\degs~declination observed on Nov. 20
(green) and Jan. 28, 2002 (red). The upper panels show brightness
temperatures for Stokes~$U$ and $Q$. In the lower panels the relative
system temperature is plotted.}
\label{bildchen_reps}
\end{figure}
Repeatability of drift scans is an important criterion for accuracy.
In Fig.~\ref{bildchen_reps} (top panels) we display two drift scans
observed at the same declination ($\delta=40$\degs). The repetition
scan was observed $10$~weeks later. Variations in the system
temperature have been corrected as described above and are shown in
the bottom panels of Fig.~\ref{bildchen_reps}. Stokes~$U$ and $Q$
values of both drift scans are in good agreement. The systematic
difference in the relative offsets is caused by a change in the
bandwidth of the IF-filters, and, as a result, the noise in all later
drift scans decreased.

\section{Correlation of the DRAO and Leiden/Dwingeloo Polarization Survey}
\begin{figure}[tb]
\centerline{\psfig{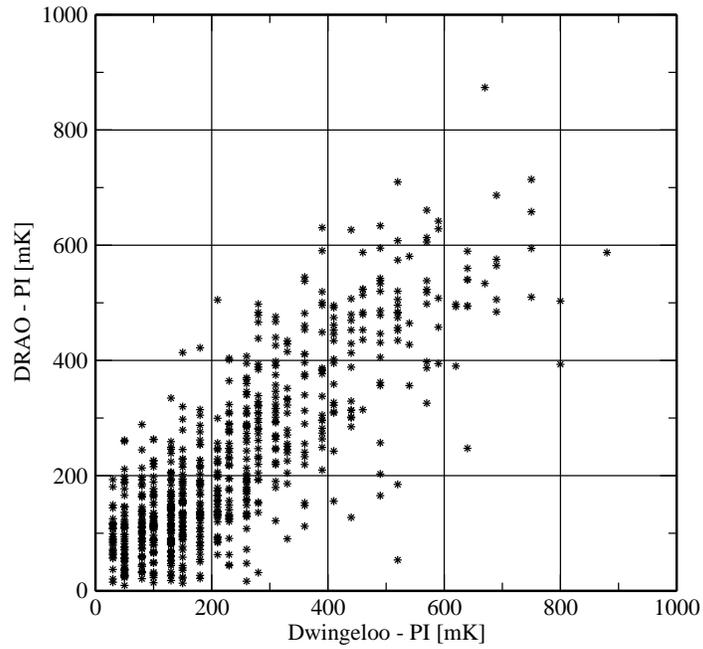}}
\caption{Comparison of polarized intensities of the DRAO and
Leiden/Dwingeloo polarization surveys.}
\label{bildchen_piall}
\end{figure}
In the previous section a comparison of both surveys is made for a
single scan at 56\degs~declination. But how do the entire data sets
compare? In Fig.~\ref{bildchen_piall} the correlation of polarized
intensities is shown for all positions which are common to both
surveys. Beside the linear correlation, a large scatter of polarized
intensities is visible. The correlation coefficients for Stokes~$U$
and $Q$ vary between 0.85 and 0.87 for this preliminary set of data.
With an average error in the Leiden/Dwingeloo data of $60$~mK, the
error in our data thus results to about $30$~mK.

\section{Preliminary Map of Polarized Intensity}
\begin{figure}[htb]
\centerline{\psfig{figure=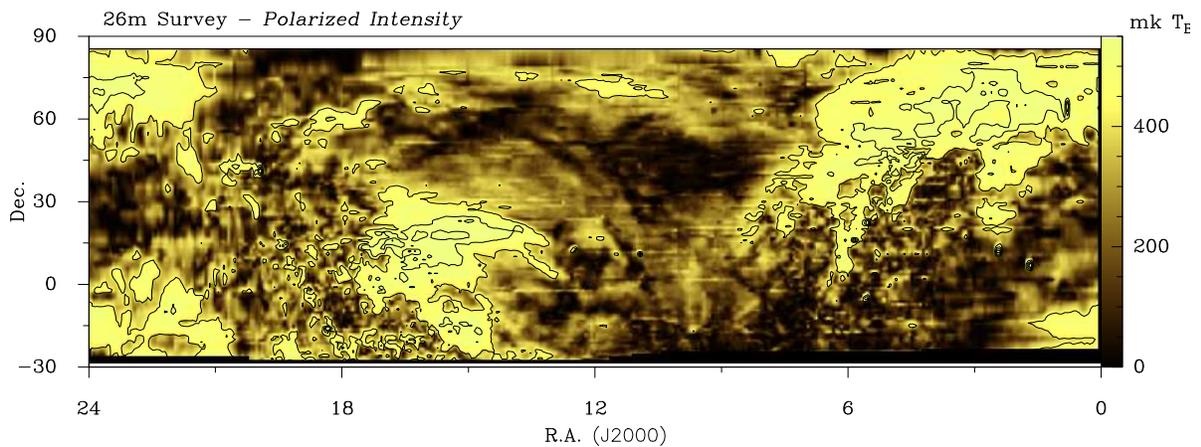,angle=270,width=15.8truecm,clip=t,bbllx=71pt,bblly=11pt,bburx=353pt,bbury=739pt}}
\caption{Preliminary map of the polarized intensity. The original data
have been convolved to match a beam size of $2$\degs. }
\label{bildchen_pi}
\end{figure}

In Fig.~\ref{bildchen_pi} we show a preliminary map of the polarized
intensity calculated by $PI=\sqrt{U^2+Q^2}$. The Stokes~$U$ and $Q$
data have been convolved to match a beam size of $4$\degs. Two bright
polarized regions of the northern sky are sticking out: the {\it
fan}-region, and the north polar spur at $\alpha=230$\degs,
$\delta=18$\degs. It is planned to continue the observations to
complete the sampling.

\section*{References}\noindent

\references

Brouw W.~N., Spoelstra, T.~A.~T. \Journal{1976}{\AaApS}{26}{129}.

Duncan A.~R., Haynes R.~F., Jones K.~L., Stewart R.~T.
\Journal{1997}{\MNRAS}{291}{279}.

Junkes N., F\"{u}rst E., Reich W. \Journal{1987}{\AaApS}{69}{451}.

Reich W. \Journal{1982}{\AaApS}{48}{210}.

Reich P., Reich W. \Journal{1986}{\AaApS}{63}{205}.

Taylor A.~R.~et al. \Journal{2003}{\AJ}{125}{3145}.

Uyan{\i}ker B.~et al. \Journal{1998}{\AaApS}{132}{401}.

\end{document}